\documentclass[twocolumn,aps,floatfix]{revtex4}
\usepackage{graphicx,epsfig,subfig,dcolumn,bm,mathrsfs,amsmath}
\usepackage{color}
\newcommand{\ud}{\mathrm{d}}

\begin{document}

\title{Elastic Properties and Line Tension of Self-Assembled Bilayer Membranes}

\author{Jianfeng Li}
\affiliation{The State Key Laboratory of Molecular Engineering of Polymers, Department of Macromolecular Science, Fudan University, Shanghai 200433, China}
\author{Kyle A. Pastor}
\affiliation{Department of Physics and Astronomy, McMaster University, 1280 Main Street West, Hamilton, Ontario, Canada L8S 4M1}
\author{An-Chang Shi}
\email[]{shi@mcmaster.ca}
\affiliation{Department of Physics and Astronomy, McMaster University, 1280 Main Street West, Hamilton, Ontario, Canada L8S 4M1}
\author{Friederike Schmid}
\affiliation{Institut f\"ur Physik, Johannes Gutenberg-Universit\"at Mainz,\\ Staudingerweg 7, D-55099 Mainz, Germany}
\author{Jiajia Zhou}
\email[]{zhou@uni-mainz.de}
\affiliation{Institut f\"ur Physik, Johannes Gutenberg-Universit\"at Mainz,\\ Staudingerweg 7, D-55099 Mainz, Germany}


\begin{abstract}
The elastic properties of a self-assembled bilayer membrane are
studied using the self-consistent field theory, applied to a model
system composed of flexible amphiphilic chains dissolved in
hydrophilic polymeric solvents. Examining the free energy of bilayer
membranes with different geometries allows us to calculate their
bending modulus, Gaussian modulus, two fourth-order membrane moduli,
and the line tension. The dependence of these parameters on the
microscopic characteristics of the amphiphilic chain, characterized
by the volume fraction of the hydrophilic component, is
systematically studied. The theoretical predictions are compared
with the results from a simple monolayer model, which approximates a
bilayer membrane by two monolayers. Finally the region of validity
of the linear elasticity theory is analyzed by examining the
higher-order contributions.
\end{abstract}


\maketitle

\section{Introduction}
\label{sec:introduction}

Amphiphilic molecules, such as lipids, surfactants and block copolymers, are composed of hydrophilic and hydrophobic components.
When dispersed in water, the amphiphilic molecules spontaneously self-assemble into a variety of structures, e.g., spherical and cylindrical micelles or bilayer membranes.
The self-assembly is driven by the competing interactions between water and different parts of the amphiphilic molecules.
Specifically, in the case of bilayers the hydrophilic parts stay on the outside of the bilayer, while the hydrophobic blocks are hidden in the interior.
At the mesoscopic scale, bilayer membranes exhibit several unique properties \cite{Lipowsky1998}: they can form closed membranes without edges, such as vesicles and cell walls; they are extremely flexible and highly deformable; and despite their flexibility, they keep their structural integrity even under strong deformations.
This combination of stability and flexibility motivates a phenomenological description of bilayer membranes, in which the membranes are modelled as two-dimensional surfaces.
The properties of the membrane within this surface model are then described by a
set of elastic parameters including the spontaneous curvature $c_0$, the bending modulus $\kappa_M$, the Gaussian modulus $\kappa_G$, and the line tension $\sigma$ of an open membrane edge.
One of the main objectives for experimentalists and theorists is to determine these elastic parameters and to understand their physical origins.
The elastic properties can be used to analyze and explain numerous phenomena associated with vesicle deformation, membrane fusion, and other relevant membrane
activities.

A number of techniques have been developed to measure the membrane elastic properties.
Particular attention has been paid to the bending modulus $\kappa_{M}$. Experimentally, $\kappa_{M}$ can be obtained either by monitoring the thermal fluctuations of a membrane \cite{Schneider1984, Faucon1989, Evans1990, Pfeiffer1993, Henriksen2004, Illya2005, Rheinstaedter2006}, or by directly measuring the force required to pull or deform a tether \cite{Bo1989, Cuvelier2005}.
Similar methods have been employed in simulation studies of the bending modulus of bilayers \cite{Goetz1999, Ayton2002, Brannigan2005, Harmandaris2006}.
Measuring the Gaussian modulus $\kappa_G$ is more challenging, since its contribution to the free energy of closed membrane systems changes only if they undergo topological changes, due to the celebrated Gauss-Bonnet theorem \cite{Millman, doCarmo, Pressley}.
A topological change often involves unstable or metastable transient membrane states.
Performing measurements in such states is a challenging task, both in experiments and simulations.
In Ref.~\cite{Siegel2004}, the Gaussian modulus of phospholipid bilayers was measured by detailed observation of the phase transition from the lamellar phase to the inverted hexagonal phase.
In Ref.~\cite{Semrau2008}, the difference between Gaussian moduli of two types of lipid bilayers was obtained.
The Gaussian modulus has also been determined in simulations by quantifying the probability that a membrane patch closes up to form a vesicle \cite{HuMingyang2012}.
Finally, a third important elastic parameter of membranes is the spontaneous curvature $c_0$.
For bilayer membranes made of two leaflets with identical composition, it is zero. We will focus on this special case in this work.

The techniques described above are designed to determine just one specific elastic parameter at a time.
An alternative strategy is to compare the free energy of membranes with different
curvatures.
By studying bilayers in different geometries (planes, cylinders, spheres), it is possible to obtain several elastic constants simultaneously.
This approach can be implemented quite naturally in theoretical studies, and there are also experimental studies based on the equilibrium between bilayer
structures with different topologies \cite{Jung2002}.
In theoretical studies, the focus has been on relating the elastic properties to
the microscopic parameters of the amphiphiles.
Based on the strong-segregation theory of grafted polymers, an analytic theory of
the monolayer elastic constants was developed by Wang and Safran \cite{WangZhen-Gang1990, WangZhen-Gang1991}.
This analytic theory was generalized to diblock copolymer bilayers by Wang
\cite{WangZhen-Gang1992}.
Similar properties for a bilayer composed of diblock copolymers were studied by Ohta and Nonomura \cite{Ohta1998}, who investigated the dependence of the bending and Gaussian moduli on the architectural parameter of the diblock
copolymers.
Apart from these analytic theories, numerical calculations using the self-consistent field theory (SCFT) have been carried out to study the elastic properties of bilayers, using diblock copolymer/homopolymer blends to mimic the amphiphile/solvent system.
Laradji and Desai \cite{Laradji1998} employed SCFT to calculate the surface tension and bending modulus from the power spectrum of capillary modes.
Matsen \cite{Matsen1999} introduced a new SCFT method to evaluate the elastic properties of a diblock monolayer.
In his work, monolayers with different surface curvature were stabilized by exerting a force on the interface.
M\"uller and Gompper \cite{Mueller2002a} extended the study to monolayers composed of pure diblock copolymers, mixtures of diblock copolymers, and triblock copolymers.
Chang and Morse \cite{Chang2006} used a pressure difference to stabilize a monolayer in curved geometries.
Nunalee \emph{et al.} \cite{Nunalee2007} presented a spontaneous curvature map for monolayers in the parameter space of the interaction parameter and the diblock architectural parameter.
In Ref.~\cite{Katsov2004}, the spontaneous curvature, bending and Gaussian moduli of a monolayer were extracted using SCFT calculations and compared to those measured in experiments.
However, despite its accuracy, the SCFT method has not yet been employed for determining the elastic properties of bilayer membranes except for the area compressibility \cite{Katsov2004}.

We turn to discussing the line tension.
Two types of line tensions can be associated with liposomes: The line tension of the phase boundary between two phases of lipids coexisting in the same membrane \cite{Baumgart2003}, and the edge free energy of an open lipid membrane.
In the present work we focus on the latter.
The line tension of an open edge is a key parameter for understanding the processes of disc-to-vesicle transformation, vesicle-pore formation and membrane fusion.
These processes often involve membranes which are highly curved in comparison to their thickness; therefore the influence of large curvatures should be examined.

Experimentally, measurements of line tensions rely on the creation of bilayer membranes with edges: One can either stabilize a bilayer membrane patch with open edges \cite{Fromherz1986}, or create pores on vesicle surfaces, or in planar membranes \cite{Zhelev1993, Karatekin2003}.
The line tension has also been determined from computer simulations of various atomistic or coarse-grained models, both for mechanically stretched membranes
\cite{Mueller1996, WangZun-Jing2005, Tolpekina2004, Wohlert2006, Otter2009} and tensionless membranes \cite{Loison2004, Jiang2004, deJoannis2006, Shinoda2011}. In comparison, theoretical studies of the line tension are scarce\cite{Moroz1997, May2000}.

In this paper, we report a unified systematic study of the elastic properties of self-assembled bilayer membranes.
Specifically, we employ SCFT formulated in different geometries to calculate the elastic moduli and line tension of a bilayer membrane.
The bilayer membrane is described by a microscopic model of flexible amphiphilic chains (AB diblock copolymers) dissolved in hydrophilic solvent molecules (A homopolymers).
Our SCFT calculations are carried out for membranes with zero or nearly zero surface tension in various geometries.
Open membranes in the form of disks and/or membranes with pores are stabilized using appropriate constraints.
The size of these stable disks or pores can be varied, which enables us to evaluate the edge line tension.
The spontaneous curvature, the bending and the Gaussian moduli can be extracted similarly to Refs.~\cite{Matsen1999, Mueller2002a, Katsov2004} by fitting the SCFT free energies of membranes with different geometries to an appropriate
energy expression for the continuum elastic model.

In the elastic model, the bilayer membrane is represented by a two-dimensional surface whose energy is described in terms of its elastic energy.
If this tensionless bilayer is not highly curved, its free energy can be well represented by a linear elastic model (the Helfrich model \cite{Helfrich1973, Ou-Yang})
\begin{equation}
  \label{eq:Helfrich}
  F = \int \left[ 2\kappa_M (M-c_0)^2 + \kappa_G G \right] \ud A + \int \sigma \ud L,
\end{equation}
where $M=(c_1+c_2)/2$ and $G=c_1 c_2$ are the local mean and Gaussian curvatures of the deformed bilayer ($c_{1,2}$ are the two principal curvatures).
The last term in Eq.\ (\ref{eq:Helfrich}) represents the edge energy of an open membrane, and it vanishes for closed membranes.
The elastic constants of interest, $c_0$, $\kappa_M$, $\kappa_G$ and $\sigma$, are the spontaneous curvature, the bending modulus, the Gaussian modulus and the edge line tension, respectively.
We emphasize that the Helfrich free energy corresponds to a linear elastic model, which includes only the lowest-order contributions to the free energy from the curvatures.

In the current work, we derive the elastic parameters from the SCFT free energy, focusing on the dependence of these quantities on the volume fraction of the hydrophilic blocks, $f_A$.
Furthermore, we carefully examine the influence of large curvatures on the accuracy of the linear elasticity theory.
We find that higher-order contributions to the membrane free energy become significant when the curvature is large.
To analyze them, we use an extension of the linear elasticity theory that includes fourth-order contributions,
\begin{eqnarray}
  F & = & \int \Big[ 2\kappa_M (M-c_0)^2 + \kappa_G G + \kappa_1 M^4 \nonumber \\
  && + \kappa_2 M^2 G + \kappa_3 G^2 \Big] \ud A  + \int \sigma \ud L,   \label{eq:Helfrich4}
\end{eqnarray}
where $\kappa_1$, $\kappa_2$ and $\kappa_3$ are fourth-order curvature moduli. With the geometries employed in our calculations, we can determine two linear combinations of the three fourth-order moduli.
Even though two such combinations do not suffice to calculate the three moduli explicitly, the values of these two constants clearly show that the higher-order terms are not negligible for highly-curved membranes.
It should be noted that in writing Eq.~(\ref{eq:Helfrich4}) we have neglected fourth-order terms involving the derivative of the curvatures, such as $(\nabla M)^2$, $\nabla^2 M^2$ and $\nabla^2 G$.
In addition, third-order terms are also not included in Eq.~(\ref{eq:Helfrich4}) due to the symmetry of the bilayers.
For a bilayer membrane composed of two identical leaflets, odd-order powers of the curvature vanish in the bending energy expansion.

The remainder of this paper is organized as follows: Section~\ref{sec:model} describes the SCFT model of bilayer membranes and the geometric constraints used in the study.
Our results on the elastic properties of the membranes are presented in section~\ref{sec:results}, and compared with the prediction from the two-monolayer approximation when appropriate.
Finally, section~\ref{sec:summary} concludes with a brief summary.

\section{Model and Method}
\label{sec:model}

In this section, we first review the theoretic framework for the calculation of the free energy of a bilayer membrane.
The general theory of SCFT has been well documented in several review articles
and monographs \cite{Schmid1998, Fredrickson2002a, Matsen2002, Shi2004_chapter, Matsen2006_chapter, Fredrickson}, and we refer readers to them for details.
We then describe the implementation of SCFT in different geometries, which we use in order to extract the elastic constants.

\subsection{Excess free energy of a bilayer membrane}
\label{sec:freeE}

The molecular model for the bilayer membranes of interest consists of a binary mixture of AB diblock (amphiphilic)
copolymers and A-type (hydrophilic) homopolymers in a volume $V$.
In the present model, the volume fraction of the hydrophilic blocks in the amphiphile is denoted by $f_A$.
For simplicity, we assume that both the copolymers and the homopolymers have equal chain length characterized by a degree of polymerization $N$.
Furthermore, we assume that the A/AB blend is incompressible, and both monomers (A and B) have the same monomer density $\rho_0$ (or the hardcore volume per monomer is $\rho_0^{-1}$) and Kuhn length $b$.
The interaction between the hydrophilic and hydrophobic monomers is described by the Flory-Huggins parameter $\chi$.
We formulate the theory in the grand-canonical ensemble and use the chemical potential of the homopolymers as a reference.
The controlling parameter is the copolymer chemical potential $\mu_c$, or its activity $z_c=\exp(\mu_c)$.
Within the mean-field approximation, the grand free energy of a binary mixture has the form \cite{Matsen1995a, 2011_cmc_bulk}:
\begin{eqnarray}
  \frac{N \mathscr{F}}{k_B T \rho_0} &=& \int
  \ud \mathbf{r} \Big[ \chi N \phi_A(\mathbf{r}) \phi_B(\mathbf{r})
  - \omega_A(\mathbf{r}) \phi_A(\mathbf{r}) \nonumber \\
  &-& \omega_B(\mathbf{r}) \phi_B(\mathbf{r}) - \xi(\mathbf{r})
  (1-\phi_A(\mathbf{r})-\phi_B(\mathbf{r})) \nonumber \\
  &-& \psi \delta(\mathbf{r}-\mathbf{r}_1) (\phi_A(\mathbf{r})
  - \phi_B(\mathbf{r})) \Big] \nonumber \\
  \label{eq:freeE}
  &-& z_c Q_{c}-Q_{h} ,
\end{eqnarray}
where $\phi_{\alpha}(\mathbf{r})$ and $\omega_{\alpha}(\mathbf{r})$ denote the local concentration and the mean field of the $\alpha$-type monomers ($\alpha=A,B$). The local pressure $\xi(\mathbf{r})$ is a Lagrange multiplier introduced to enforce incompressibility.
A second Lagrange multiplier, $\psi$, is used to implement constraints for stabilizing the bilayer in different geometries.
A delta function, $\delta(\mathbf{r}-\mathbf{r}_1)$, is used to ensure that the $\psi$ field only operates on the interface at a prescribed position $\mathbf{r}_1$.
The last two terms in Eq.~(\ref{eq:freeE}) are the configurational entropies, related to the single-chain partition functions for two types of polymer, $Q_{c}$ and $Q_{h}$.
For the copolymer, the partition function has the form $Q_{c} = \int \ud \mathbf{r} q_{c}(\mathbf{r},1)$, where $q_{c}(\mathbf{r},s)$ is an end-integrated propagator, and $s$ is a parameter that runs from 0 to 1 along the length of the polymer.
The propagator satisfies the modified diffusion equation
\begin{equation}
  \label{eq:mde}
  \frac{\partial}{\partial s} q_{c}(\mathbf{r},s) = R_g^2 \nabla^2 q_{c}(\mathbf{r},s) - \omega_{\alpha}(\mathbf{r}) q_{c}(\mathbf{r},s),
\end{equation}
where $R_g=b\sqrt{N/6}$ is the gyration radius of the copolymer
chains.
The mean field $\omega_{\alpha}(\mathbf{r})$ is a piece-wise function where $\alpha=A$ if $0<s<f_A$ and $\alpha=B$ if $f_A<s<1$.
The initial condition is $q_c(\mathbf{r},0)=1$. Since the copolymer has two distinct ends, a complementary propagator $q_c^{\dagger}(\mathbf{r},s)$ is introduced.
It satisfies Eq.~(\ref{eq:mde}) with the right-hand side multiplied by $-1$, and the initial condition $q_c^{\dagger}(\mathbf{r},1)=1$.
For the homopolymer, one propagator $q_h(\mathbf{r},s)$ is sufficient, and the single-chain partition function has the form $Q_h = \int \ud \mathbf{r} q_h(\mathbf{r},1)$.
In our calculations the modified diffusion equations are solved in real space using the Crank-Nicolson method \cite{NR3}.

The SCFT method employs a mean field approximation to evaluate the free energy using a saddle-point technique.
This demands the functional derivatives of the expression (\ref{eq:freeE}) to be zero,
\begin{equation}
  \frac{\delta \mathscr{F}}{\delta\phi_{\alpha}} = \frac{\delta \mathscr{F}}{\delta\omega_{\alpha}} = \frac{\delta \mathscr{F}}{\delta\xi} = \frac{\delta \mathscr{F}}{\delta\psi} =0.
\end{equation}
Carrying out these functional derivatives leads to the following mean-field equations:
\begin{eqnarray}
  \phi_A(\mathbf{r}) &=& \int_0^1 \ud s \, q_h(\mathbf{r},s) q_h(\mathbf{r},1-s) \nonumber \\
  && + z_c \int_0^{f_A} \ud s \, q_c(\mathbf{r},s) q_c^{\dagger}(\mathbf{r},s), \\
  \phi_B(\mathbf{r}) &=&  z_c \int_{f_A}^1 \ud s \, q_c(\mathbf{r},s) q_c^{\dagger}(\mathbf{r},s), \\
  \omega_A(\mathbf{r}) &=& \chi N \phi_B(\mathbf{r}) + \xi(\mathbf{r}) - \psi \delta(\mathbf{r}-\mathbf{r}_1),\\
  \omega_B(\mathbf{r}) &=& \chi N \phi_A(\mathbf{r}) + \xi(\mathbf{r}) + \psi \delta(\mathbf{r}-\mathbf{r}_1),\\
  1 &=& \phi_A(\mathbf{r})+\phi_B(\mathbf{r}),\\
  \phi_A(\mathbf{r}_1) &=& \phi_B(\mathbf{r}_1).
\end{eqnarray}
These equations can be solved by iteration.

We are interested in the free energy of a system containing a bilayer membrane compared to that of a reference system without a bilayer, \emph{i.e.}, a homogeneous blend of copolymers and homopolymers.
The free energy for the homogeneous bulk phase can be computed analytically,
\begin{eqnarray}
  \frac{N\mathscr{F}_{\rm bulk}}{k_B T \rho_0 V} &=& (1-\phi_{\rm bulk}) \big[\ln(1-\phi_{\rm bulk})-1 \big] \nonumber \\
  &+& \phi_{\rm bulk} \big[\ln \phi_{\rm bulk} - 1 \big] \nonumber \\
  &+& \chi N (1-f_A) \phi_{\rm bulk} \big[ 1 - \phi_{\rm bulk} + f_A\phi_{\rm bulk} \big] \nonumber \\
  &-& \mu_c \phi_{\rm bulk},
\end{eqnarray}
where $\phi_{\rm bulk}$ is the bulk copolymer concentration.
In the grand canonical ensemble, the bulk copolymer concentration depends on the copolymer chemical potential $\mu_c$ {\em via} the relation:
\begin{eqnarray}
  \mu_c &=& \ln \phi_{\rm bulk} - \ln ( 1- \phi_{\rm bulk}) \nonumber \\
  &+& \chi N (1-f_A) \big[ 1-2(1-f_A)\phi_{\rm bulk} \big].
\end{eqnarray}

For a bilayer membrane, its excess free energy ($\mathscr{F}-\mathscr{F}_{\rm bulk}$) is proportional to the area of the membrane.
We can define an excess free energy density, $F$, as the free energy difference between the systems with and without the bilayer membrane, divided by the area, $A$, of the membrane,
\begin{equation}
  \label{eq:F_excess}
  F = \frac{N(\mathscr{F}-\mathscr{F}_{\rm bulk})}{k_B T \rho_0 A}.
\end{equation}
Another useful quantity characterizing the membrane is the
copolymer excess per unit area,
\begin{equation}
  \label{eq:omega}
  \Omega = \frac{1}{A} \int \ud \mathbf{r} \Big[ \phi_c(\mathbf{r}) - \phi_{\rm bulk} \Big],
\end{equation}
where $\phi_c(\mathbf{r})$ is the local concentration of the AB
diblock copolymers.

\subsection{Geometrical constraints}
\label{sec:geometry}

In order to extract information on the various elastic properties, we calculate the excess free energy of a bilayer membrane in the following five geometries: (i) an infinite planar bilayer membrane, (ii) a cylindrical bilayer membrane with a radius $r$, which is extended to infinity in the axial direction, (iii) a spherical bilayer with a radius $r$, (iv) an axially symmetric disk-like membrane patch with a radius $R$, and (v) a planar membrane with a circular pore of radius $R$.
The first three geometries, which can be reduced to a one-dimensional problem by an appropriate coordinate transformation, are employed to extract the bending modulus and the Gaussian modulus as well as higher-order curvature moduli.
The last two geometries, which are two-dimensional systems due to their axial
symmetry, are used to extract the line tension of the membrane edge.

To stabilize a bilayer membrane in these geometries, a constraint term, $\psi\delta(\mathbf{r}-\mathbf{r}_1)(\phi_A - \phi_B)$, has been included in the free energy expression (\ref{eq:freeE}).
From a mathematical point of view, this term ensures that at the point specified by $\mathbf{r}_1$, the concentrations of the hydrophilic and hydrophobic monomers are equal, \emph{i.e.}, $\phi_A(\mathbf{r}_1) = \phi_B(\mathbf{r}_1)$.
In the cylindrical and spherical geometries, this constraint sets the curvature radius of the membrane, and in the disk/pore geometry, it defines the size of the disk/pore.
It should be noted that in the cylindrical and spherical case, the constraint is applied to the outer monolayer only, and the inner monolayer is free.
This allows the bilayer to optimize its thickness.
We found that constraining the inner monolayer while removing the constraint from the outer monolayer lead to identical results.
In order to ensure that the calculated membrane properties are not affected by finite size effects, we choose a calculation box large enough such that the monomer densities reach their bulk values at the boundaries.
With these geometric constraints, we obtained the excess free energies for the
five geometries, which are denoted $F^{X}$, where $X=0,C,S,D,P$ for the planar, the cylindrical, the spherical, the disk and the pore geometry, respectively. Furthermore, we focus on tensionless membranes in this work.
The chemical potential of the amphiphilic diblock copolymers, $\mu_c$, is adjusted carefully such that the grand free energy of the bulk system and of a planar bilayer are identical, \emph{i.e.}, $F^0 = 0$.

For a curved bilayer of finite thickness, the definition of the interface position or the membrane surface involves a certain degree of arbitrariness.
To be consistent with our constraint method, we define the bilayer interface to be at the mid point of the two positions where A- and B-segment concentrations are equal.
For a cylindrical membrane with radius $r$, the mean curvature is $M=1/(2r)$ and the Gaussian curvature vanishes.
For a spherical membrane with radius $r$, the mean curvature and the Gaussian
curvature are $M=1/r$ and $G=1/r^2$, respectively.
With these curvatures, the modified Helfrich free energy (\ref{eq:Helfrich4}) can be written in terms of the curvature $c=1/r$ in the cylindrical and spherical geometries,
\begin{eqnarray}
  \label{eq:fC_fit}
  F^C &=& -2 \kappa_M c_0 c + \frac{\kappa_M}{2} c^2 + B_C c^4, \\
  \label{eq:fS_fit}
  F^S &=& -4 \kappa_M c_0 c + (2\kappa_M + \kappa_G) c^2 + B_S c^4,
\end{eqnarray}
where the higher order parameters $B_C=\kappa_1/16$ and $B_S=\kappa_1+\kappa_2+\kappa_3$ will be termed fourth-order cylindrical and spherical modulus, respectively. In the future studies, one can use
ellipsoidal geometry with revolutionary symmetry to
extract these moduli separately, where the entire
shape of the outer layer has to be fixed during the SCFT calculations.
Following Ref.~\cite{Katsov2004}, we use the thickness, $d=4.3R_g$, of a planar
bilayer with equal fractions of hydrophilic and hydrophobic segments ($f_{A}=0.50$) as the unit of length, where the thickness in planar geometry is given by the copolymer excess (\ref{eq:omega}), $d=\Omega$.
For the surface tension, a natural unit is the interfacial free energy per unit area between coexisting A and B homopolymers in the limit of large $\chi N$, $\gamma_{\rm int}=\sqrt{\chi N/6} \,\rho_0 k_BTb/\sqrt{N}$.
Using this convention, the modulus can be made dimensionless:
\begin{equation}
  \tilde{\kappa}_M = \frac{\kappa_M}{\gamma_{\rm int} d^2}, \quad
  \tilde{\kappa}_G = \frac{\kappa_G}{\gamma_{\rm int} d^2},
\end{equation}
\begin{equation}
  \label{eq:BCS}
  \tilde{B}_C = \frac{\kappa_1}{16 \gamma_{\rm int} d^4}, \quad
  \tilde{B}_S = \frac{\kappa_1+\kappa_2+\kappa_3}{\gamma_{\rm int} d^4}.
\end{equation}

Similar to the procedures described above, the line tension of an open edge can be calculated by examining the free energy of membranes as a function of the disk or pore sizes,
\begin{eqnarray}
  \label{eq:fd}
  F^D A &=& 2\pi \sigma R + {\rm const}, \\
  \label{eq:fp}
  F^P A &=& 2\pi \sigma R + {\rm const},
\end{eqnarray}
where $R$ is the diameter of the disk/pore.
The definition of the edge position is again somewhat arbitrary for a membrane of finite thickness.
However, the resulting value for $\sigma$ does not depend on the specific convention used for calculating $R$.
Changing the convention will only affect the constant offset in the free energy expressions, Eqs.~(\ref{eq:fd}) and (\ref{eq:fp}).
Here, we define $2R$ as the distance between two diametrally opposed points in
the bilayer mid-plane that satisfy the constraint $\phi_A(\mathbf{r}) = \phi_B(\mathbf{r})$.
Using the length unit defined above, we define a unit for the line tension, $\sigma_0=k_BT \rho_0 d^2/N$.
The dimensionless line tension is then given by
\begin{equation}
  \tilde{\sigma} = \frac{\sigma}{\sigma_0}.
\end{equation}


\section{Results and Discussion}
\label{sec:results}

In this section, we first present the SCFT results for the second-order elastic moduli ($\kappa_M$ and $\kappa_G$) and discuss their relation to the microscopic parameter $f_A$.
We then examine highly curved bilayer membranes and demonstrate that the contribution to the free energy from the higher-order moduli can be significant.
Finally we study the bilayer membrane in disk/pore geometries and compute the line tension by varying the structure size.
Since we mainly focus on the effect of the hydrophilic volume fraction, $f_A$, on the bilayer properties, most of the results are presented for a particular interaction strength between the A and B monomers.
That is, we have fixed $\chi N=30$ unless otherwise stated.
For this intermediate segregation case, the chemical potential of the copolymer is around $\mu=4.6k_BT$ when a planar bilayer membrane becomes tensionless.
It should be noted that the interaction parameter is also important, but a change in the interaction parameter does not produce any qualitative changes to
the $f_{A}$-dependence of the elastic moduli.

\subsection{Linear Elasticity: Bending and Gaussian moduli}

\begin{figure}[htp]
  \includegraphics[width=1.0\columnwidth]{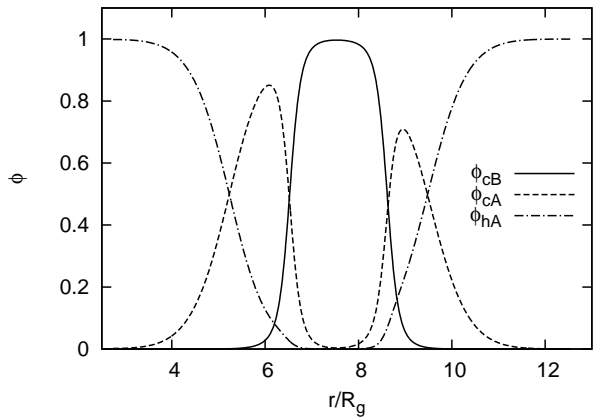}
  \caption{The concentration profile for a bilayer membrane in the spherical geometry with a radius $R=7.7R_g$ and $f_A=0.50$. The concentration for the hydrophilic solvent ($\phi_{hA}$), the hydrophilic ($\phi_{cA}$) and hydrophobic ($\phi_{cB}$) segments of the amphiphiles, are shown as dot-dash, dashed and solid lines, respectively.}
  \label{fig:profile}
\end{figure}

From the SCFT free energy of cylindrical and spherical membranes at large radii or small curvatures, the second-order moduli, the bending modulus $\kappa_M$ and the Gaussian modulus $\kappa_G$, can be calculated by fitting the free energy curve to the Helfrich model.
The basic structural properties of a self-assembled bilayer membrane are the spatial distribution of the hydrophilic and hydrophobic monomers across it.
A typical concentration profile for a bilayer membrane in the spherical geometry with a radius $R=7.7R_g$ and $f_A=0.50$ is shown in Fig.~\ref{fig:profile}.
Within the spherical bilayer membrane, since the area of the inner interface is smaller than the outer one, the hydrophilic monomers (A-blocks) in the inner leaflet have to pack more closely and have a higher local concentration in comparison to the outer leaflet.
At the same time, the hydrophilic monomers in the inner interface also have a wider distribution in the axial direction, which indicates that the hydrophilic chains in the same region are more stretched than the outer chains.
The loss of conformational entropy due to the local extension of hydrophilic chains is the main contribution to the bending energy of the membranes.

\begin{figure}[htp]
   \centering
   \includegraphics[width=1.0\columnwidth]{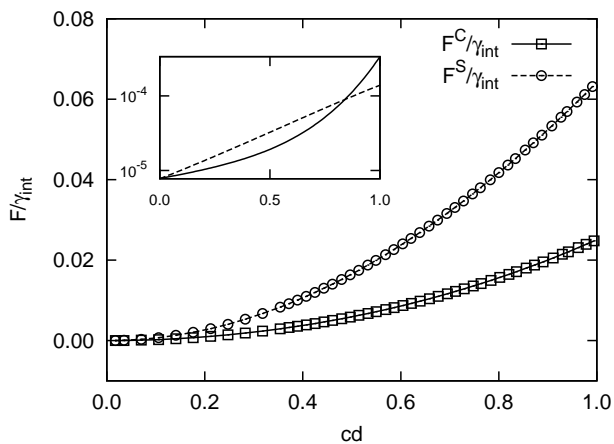}
   \caption{The excess free energy for a tensionless bilayer membrane with
   $f_A=0.50$ as a function of the dimensionless curvature $cd$ in the
   cylindrical (square symbols) and spherical (sphere symbols) geometries.
   The curvature of the membrane is maintained by fixing the position of the
   outer monolayer. Alternatively, one can also fix the inner monolayer.
   The difference of the free energy obtained with these different implementations
   of the constraints is shown in the inset for the cylindrical (solid line)
   and spherical geometry (dashed line).}
   \label{fig:freeE}
 \end{figure}

The excess free energies for a tensionless bilayer with $f_A=0.50$ in the cylindrical and spherical geometries are shown in Fig.~\ref{fig:freeE}.
The free energies are plotted as a function of the dimensionless curvature $cd$. In the inset, we show the difference between the free energies obtained by different methods of applying the constraint.
For highly curved membranes, the free energy with fixed inner monolayer is slightly higher than that with fixed outer monolayer.
In general, one would expect that the free energy will depend on the specifics of how the constraint is applied, with the exceptions of (meta)stable states and saddle points where the conjugated field $\psi$ vanishes.
However, for the tensionless bilayer membranes studied here, the deviation introduced by using different constraints is small, so we opt for the method of
applying the constraint to the outer monolayer in this study.

Ideally, one can fit the excess free energy as a polynomial function of the curvature up to fourth order, then extract the elastic properties from the fitting parameters using Eqs.~(\ref{eq:fC_fit}) and (\ref{eq:fS_fit}).
The zeroth-order term should be zero for a tensionless membrane, and indeed this is the case as shown in Fig.~\ref{fig:freeE}, where the excess free energies go to zero when the curvature goes to zero.
The linear term in the free energy plot is associated with the spontaneous curvature of the membranes.
For a bilayer membrane consisting of two identical leaflets, the spontaneous curvature is zero by symmetry.
This can be seen from Fig.~\ref{fig:freeE} where the slope of the free energy curves also vanishes at the zero curvature.
Furthermore, the symmetry of the membranes dictates that the higher odd-order terms also vanish.

\begin{figure}[htp]
  \includegraphics[width=1.0\columnwidth]{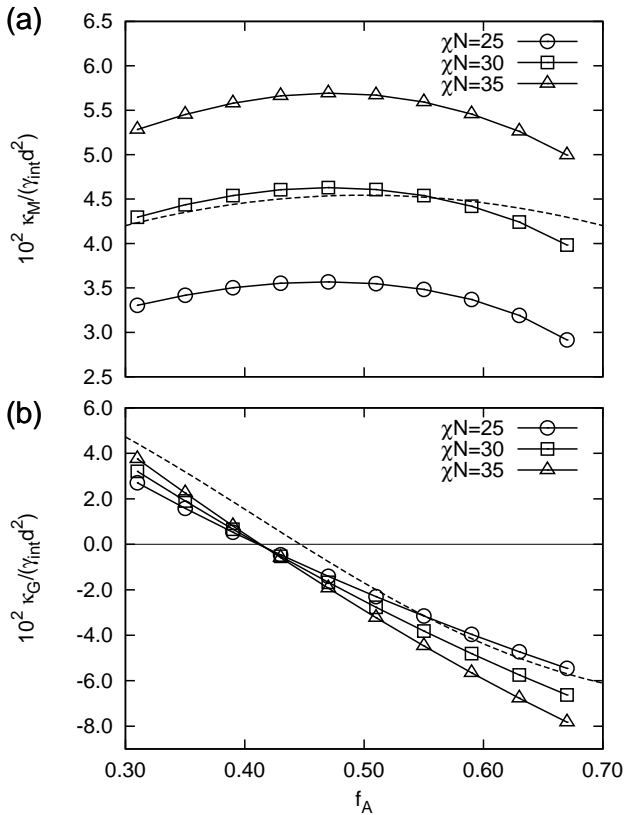}
  \caption{The bending modulus $\kappa_M$ and the Gaussian modulus $\kappa_G$
  of a bilayer membrane as a function of the hydrophilic fraction $f_A$.
  The results for different interaction strength $\chi N$ are shown in
  different symbols. We note that the reference energies,
  $\gamma_{\rm int}$, increase with $\chi N$ via
  $\gamma_{\rm int} \propto \sqrt{\chi N}$. This trivial dependence
  has been eliminated by plotting the dimensionless renormalized quantities.
  The dashed curves correspond to the results from the two-monolayer
  approximation, Eqs.\ (\ref{eq:couple_kM}) and (\ref{eq:couple_kG}),
  for $\chi N=30$.}
  \label{fig:kappa}
\end{figure}

The first non-zero coefficient in the curvature expansion of the membrane free energy is the second-order term, which is related to the bending and Gaussian moduli.
Fig.~\ref{fig:kappa} shows $\kappa_M$ and $\kappa_G$, as a function of the hydrophilic volume function $f_A$.
Beyond $f_A \ge1-\sqrt{2/\chi N}$ ($=0.74$ for $\chi N=30$), bilayer membranes are unstable with respect to the homogeneous phase.
In the stable regime below this value, the dependence on $f_A$ is quite different for the two quantities, $\kappa_M$ and $\kappa_G$.
The bending modulus, $\kappa_M$, on the one hand, is not very sensitive to the amphiphilic architecture specified by $f_{A}$. As one might expect, it exhibits a weak maximum close to $f_A=0.50$ (symmetric amphiphiles), which is however not exactly symmetric with respect to $f_A=0.50$ due to the different solubility of the amphiphilic molecules.
Molecules with longer hydrophilic parts can be dissolved in the solvent more easily, resulting in a higher excess at the interface.
The Gaussian modulus, $\kappa_G$, on the other hand, is a strongly varying
monotonically decreasing function of the hydrophilic volume fraction, and its value changes from positive to negative, crossing zero at around $f_A=0.41$.
The planar bilayer becomes unstable for positive values of $\kappa_G$.
It is interesting to note that the ratio of the Gaussian modulus and the bending modulus, $\kappa_{G}/\kappa_{M}$, decreases from roughly 1 to -2 as $f_{A}$
is increased, changing its sign at about $f_A=0.41$.

The effect of varying the chain interaction strength on the qualitative behavior of the second-order moduli is small, as shown in Fig~\ref{fig:kappa} for different values of $\chi N$.
The bending modulus $\kappa_M$ increases with increasing interaction parameter
$\chi N$.
The overall dependence on the hydrophilic architecture $f_A$ is the same for different $\chi N$.
Even more interestingly, the Gaussian modulus $\kappa_G$ shows a similar relation.
When the interaction strength increases, the magnitude of the modulus also increases, and when the interaction strength decreases, the magnitude of $\kappa_G$ does the same.
The zero crossing point for all interaction strengths is the same at approximately $f_A=0.41$.
The architectural dependence is not drastically affected as the Gaussian modulus $\kappa_G$ is still a monotonically decreasing function in all cases.
This result shows that the choice of interaction parameter has no qualitative effect on the behaviour of the elastic moduli as a function of the amphiphile architecture.

The calculated elastic properties of a bilayer membrane can be compared with those of its corresponding monolayers.
The monolayer properties have been well studied with the SCFT by several authors
\cite{Matsen1999, Mueller2002a, Katsov2004, Chang2006, Nunalee2007}.
The bilayer properties can be derived from monolayer parameters by a two-monolayer approximation as discussed in Appendix~\ref{app:moduli}.
In this simple model the bilayer membrane is depicted by two independent monolayers sticking together without any specific interactions, and the elastic properties of the bilayer can be expressed in terms of the monolayer counterparts. For instance, for a symmetric bilayer \cite{Petrov1984, Siegel2004, Marsh2006} we
have,
\begin{eqnarray}
  \label{eq:couple_kM}
  \kappa_M &=& 2 \bar{\kappa}_M , \\
  \label{eq:couple_kG}
  \kappa_G &=& 2 \big[ \bar{\kappa}_G - 4\bar{\kappa}_M \bar{c}_0 \delta \big].
\end{eqnarray}
In the above expressions, over-barred quantities denote the properties of the monolayer, and $\delta$ is the distance between the midplane of the bilayer and the monolayer interface.
The value of $\delta$ is a linearly decreasing function of $f_A$ (not shown
here).
We also calculated the monolayer properties with the SCFT and the results are similar to those reported in Ref.~\cite{Katsov2004}.
The spontaneous curvature of the monolayer increases in an approximately linear manner with the hydrophilic fraction $f_A$, with $\bar{c}_0=0$ for symmetric amphiphiles with $f_A=0.50$.
In Fig.~\ref{fig:kappa}, the prediction from the monolayer properties are plotted as dashed curves for $\chi N=30.0$.
It is interesting to notice that the simple two-monolayer model is in good qualitative agreement with the exact results from SCFT.

The bending modulus of the bilayer is roughly twice that of the monolayer.
In previous simulations and experiments, stacking monolayers are also found to reinforce the rigidity of the membrane \cite{Siegel2004, Kurtisovski2007}.
The prediction for the bending modulus from the two-monolayer approximation must necessarily be symmetric about $f_A=0.50$, due to the fact that the bending modulus
for monolayer is symmetric about $f_A=0.50$.
The SCFT results are slightly different, with the maximum shifted to a smaller
hydrophilic fraction.
This can be attributed to the difference between the two-monolayer approximation and the full bilayer calculation: In the latter case one considers two apposing (and possibly coupled) monolayers, while in the former case, the apposing monolayer is replaced by a hydrophobic homopolymer melt.
In our case, where copolymers and homopolymers have equal length, one would expect the monolayer approximation to be better when the composition of copolymers approaches that of hydrophobic (B-)homopolymers (small $f_A$).

The Gaussian modulus for a bilayer involves not only the two curvature moduli of the monolayer, but also its spontaneous curvature, $\bar{c}_0$.
At $\bar{c}_0 = 0$, the Gaussian modulus is slightly negative.
Due to the interplay of monolayer stiffness and spontaneous curvature, it is further reduced at $f_A > 0.50$, and shifted upwards for $f_A < 0.50$.

\subsection{Nonlinear Elasticity: Higher-order Moduli}

In the above subsection, the SCFT results demonstrate that the Helfrich model gives an excellent description of the free energy of slightly curved bilayer membranes.
The Helfrich expression essentially corresponds to a linear elasticity theory of membranes, where the higher-order contributions are ignored.
It is therefore desirable to examine the region of validity of the linear elastic
theory.
Specifically, we expect that the higher-order terms in Eq.~(\ref{eq:Helfrich4}) can no longer be ignored in systems of highly curved membranes.
Examining the relative contribution of the higher-order terms should then allow us to establish regions of validity for the Helfrich model.
In Ref.~\cite{Harmandaris2006}, Harmandaris and Deserno employed a series of mesoscopic simulations of a coarse-grained model to test the validity of the Helfrich model and found that the quadratic Helfrich energy remains valid for
membranes with curvature radius comparable to the bilayer thickness.
However, it should be noted that these authors employed the Cooke model \cite{Cooke2005, Cooke2005a}, which replaces the lipid tail by two beads.
This high level of coarse-graining may limit the ability of the model to describe highly curved membrane structures.
In more refined models, one important contribution to higher-order terms is the change of the packing of the lipid tails upon bending the bilayer.
In Ref.~\cite{Risselada2011}, Risselada \emph{et al.} used the {\sc MARTINI} model to study vesicles formed of lipid bilayers.
The higher-order contribution to the Helfrich Hamiltonian is important for explaining the inverted domain sorting under uniaxial compression.
The higher-order elastic constants are also shown to stabilize the intermediate states of open vesicles in the disk-to-vesicle transition \cite{LiJianfeng2013a}.
In the SCFT calculations, the stretching of the amphiphilic chains is captured, which allows us to examine the effects of the highly curved membranes.

\begin{figure}[htp]
  \includegraphics[width=1.0\columnwidth]{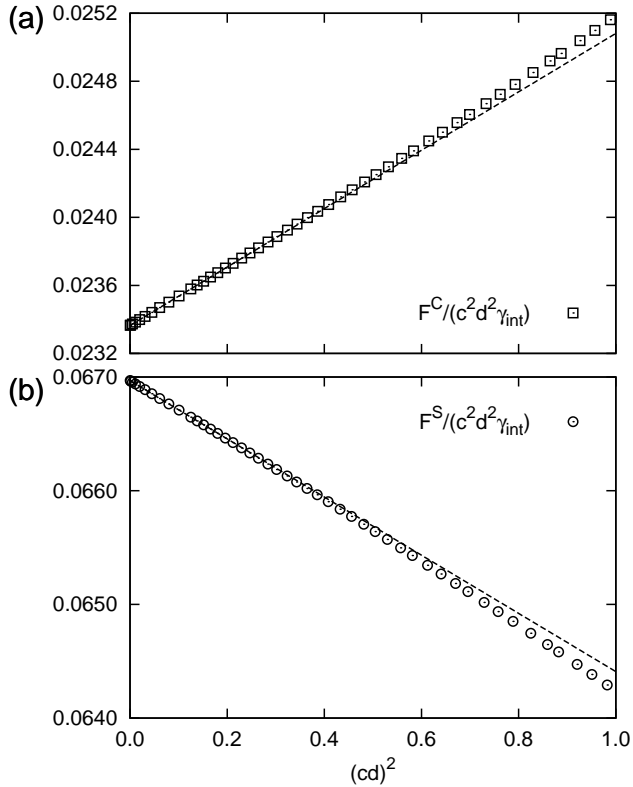}
  \caption{Rescaled free energy curves $F/(cd)^2$ plotted as a function of
  squared curvature $(cd)^2$ for (a) the cylindrical and (b) the spherical
  membranes. Symbols and parameters are the same as in Fig.~\ref{fig:freeE}. }
  \label{fig:freeE_m}
\end{figure}

\begin{figure}[htp]
  \includegraphics[width=1.0\columnwidth]{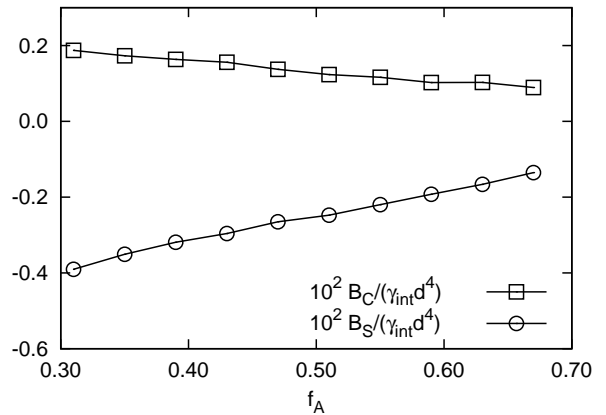}
  \caption{Fourth-order moduli $B_C$ and $B_S$ as a function
  of the hydrophilic fraction $f_A$.}
  \label{fig:BCS}
\end{figure}

The effect of the higher-order terms on the membrane free energy can be demonstrated by plotting the rescaled free energy, $F/(cd)^2$ , as a function of the effective curvature $(cd)^2$, as shown in Fig.~\ref{fig:freeE_m} for both cylindrical and spherical geometries.
If the quadratic expression of the Helfrich model is strictly valid, the scaled free energy $F/(cd)^2$ should remain constant in these plots.
Contrary to this expectation, the SCFT results shown in Fig.~\ref{fig:freeE_m}  reveal that the scaled free energy of the membrane changes as $(cd)^2$ is increased.
For small values of $(cd)^2$, the scaled free energy is approximately a linear function of $(cd)^2$. The slope of this linear function gives the quartic correction to the free energy.
For highly curved bilayers, contributions of terms with even higher order come into play, and the scaled free energy deviates from the straight line.
From these observations we conclude that higher order terms should be included when describing the behavior of highly curved membranes within an elastic model. Quantitatively, the fourth-order term can contribute as much as $10\%$ to the total free energy for highly curved membranes.
This contribution should not be neglected.
Due to the limited number of geometries employed in our calculations, we are only able to calculate the cylindrical and spherical fourth order moduli defined in Eq.~(\ref{eq:BCS}).
In Fig.~\ref{fig:BCS}, these two higher-order moduli are plotted as a function of the hydrophilic fraction $f_A$.
The general trend is that the higher-order contribution is positive for cylindrical
geometry, but negative for spherical bilayers.

\begin{figure}
  \includegraphics[width=1.0\columnwidth]{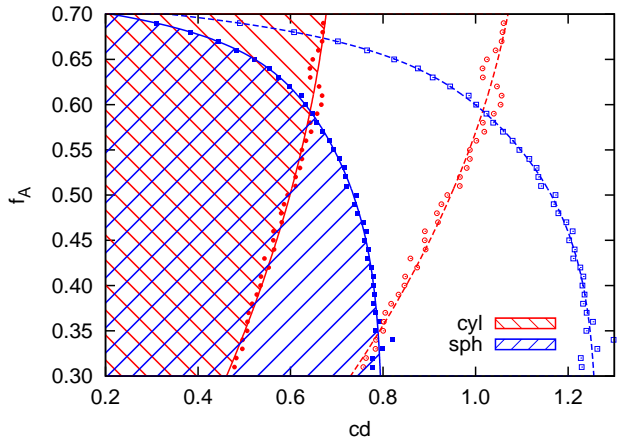}
  \caption{Curvature boundaries for both the cylindrical and spherical
  bilayer geometries. The results for the criteria $\Delta E^*=2\%$
  are shown by filled symbols and solid lines, and the results for
  $\Delta E^*=5\%$ are shown by unfilled symbols and dashed lines.
  Regions where the second-order terms adequately describe the
  bending energy (with $\Delta E^*=2\%$) for both geometries are
  filled with crossed stripes. Unpatterned regions or regions
  filled with simple stripes correspond to curvatures where fourth-order
  corrections are needed to describe the bending energy for at least
  one geometry (with $\Delta E^*=2\%$). }
  \label{fig:boundary}
\end{figure}

With our results for the second and fourth-order moduli, we can now determine a boundary separating the region where the fourth-order corrections are negligible from that where the fourth-order terms start to play an important role.
We require that the relative energy difference,
\begin{equation}
 \Delta E = \frac{|F_{4}^{C}-F_{2}^{C}|}{F_{2}^{C}},
\end{equation}
does not exceed a given threshold $\Delta E^*$.
Here $F_{4}^{C}$ is the free energy for the cylindrical geometry including the
fourth-order contributions to the free energy, and $F_{2}^{C}$ is the free energy of the cylindrical geometry assuming that the fourth-order moduli are negligible.
A similar definition is used for the spherical geometry.
Applying Eqs.~(\ref{eq:fC_fit}) and (\ref{eq:fS_fit}), one obtains the following expressions for the critical curvatures for the cylindrical and spherical bilayers:
\begin{eqnarray}
 c^*_{C}&=&\sqrt{\frac{\kappa_M \Delta E^*}{2 B_{C}}} \\
 c^*_{S}&=&\sqrt{\frac{(2 \kappa_M + \kappa_G)\Delta E^*}{B_{S}}}.
\end{eqnarray}
Using the values of the moduli as a function of hydrophilic chain fraction ($f_A$) as obtained in the previous sections, the boundary of validity for the linear elastic theory is determined for both the cylindrical and spherical geometries.
The results for $\Delta E^*=2\%$ and $5\%$ are shown in Fig.~\ref{fig:boundary}.

Let us first examine the criterion $\Delta E^*=2\%$, which corresponds in Fig.~\ref{fig:boundary} to the filled symbols and solid lines.
In the patterned regions the second-order description of the bilayer bending energy is valid up to the prescribed accuracy $\Delta E^*$.
In the unpatterned regions, this description becomes less accurate and the fourth-order moduli must be included to accurately describe the membrane energy.
Above a hydrophilic chain fraction of $f_A\approx0.58$ the absolute boundary on the critical curvature is dominated by the spherical geometry.
In this region, membranes in cylindrical geometry can be described by the second-order bending moduli for a larger span of curvatures up to $cd=0.70$ for a chain fraction of approximately $f_A=0.70$, but for spherical geometry, the valid region is significantly reduced to about $cd=0.2$ for $f_A=0.70$.
This trend is reversed below $f_A\approx0.58$ where the absolute boundary is set by the cylindrical geometry.
In this range of chain fractions, the spherical corrections are relatively less
important, which implies that the second-order descriptions can be used up to curvatures of $cd \approx 0.8$ for a chain fraction length of $f_A=0.30$.
The unpatterned regions corresponds to parameter sets in the ($cd,f_A$)-plane where the fourth-order terms must be included in order to describe the bending energy of the bilayer up to the accuracy $\Delta E^*=2\%$, regardless of membrane geometry.
If the accuracy criterion is relaxed to $\Delta E^*=5\%$, we obtain the curves shown in Fig.~\ref{fig:boundary} in unfilled symbols and dashed lines, and the valid region for the second-order description is extended to larger curvatures up to around $cd \sim 1$.

\subsection{Line Tension of Membrane Edge}

\begin{figure}[htp]
  \includegraphics[width=1.0\columnwidth]{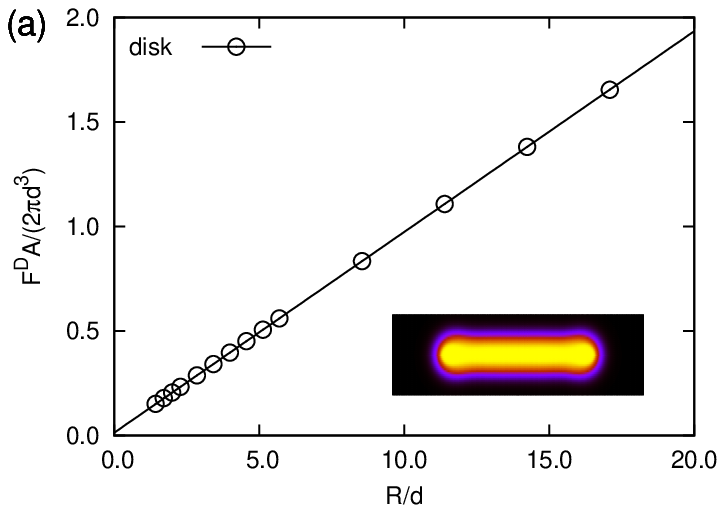}
  \includegraphics[width=1.0\columnwidth]{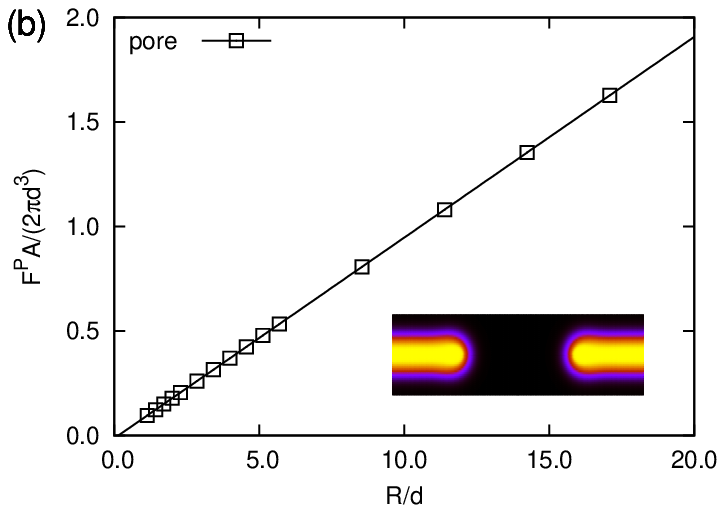}
  \caption{The excess free energy for a disk-like bilayer membrane (a), and a pore in a planar bilayer membrane (b). The amphiphilic molecules are symmetric, $f_A=0.50$. The insets show representative density
profiles for the amphiphilic molecules of the corresponding structures.}
  \label{fig:line}
\end{figure}

The calculation for the line tension of the membrane edge must be performed in two dimensions.
In our study, two different geometries, a circular disk and a circular pore, are employed.
The excess free energies of the tensionless disk-like membranes of different size
are calculated using the SCFT, and the results are shown in Fig.~\ref{fig:line}(a).
The free energy is a linear function of the radius $R$.
The slope of the curve is then used to obtain the line tension.
A typical density profile of a disk-like membrane is shown in the inset of Fig.~\ref{fig:line}(a).
The cross-section contour of the disk can be approximated by two semicircles being separated by two parallel lines.
Closely examining this contour, we note that the thickness in the middle of the disk is slightly smaller than the diameter of the semicircle, and each semicircle is connected  smoothly to the two parallel lines by two inverted arcs.

In Fig.~\ref{fig:line}(b), we plot the excess free energy for a bilayer membrane with a pore of varying radius.
Similar to the case of the disk structure, the free energy is a linear function of the radius $R$.
Similar curves had been obtained in Ref.~\cite{Katsov2006}, and in Ref.~\cite{Wohlert2006}, where the free energy at very small radius was found to deviate from the linear dependence.
A cross-section of the pore structure is shown in the inset of Fig.~\ref{fig:line}(b).
The thickness near the edge is slightly larger than that away from the edge, similar to the disk geometry.

\begin{figure}[htp]
  \includegraphics[width=1.0\columnwidth]{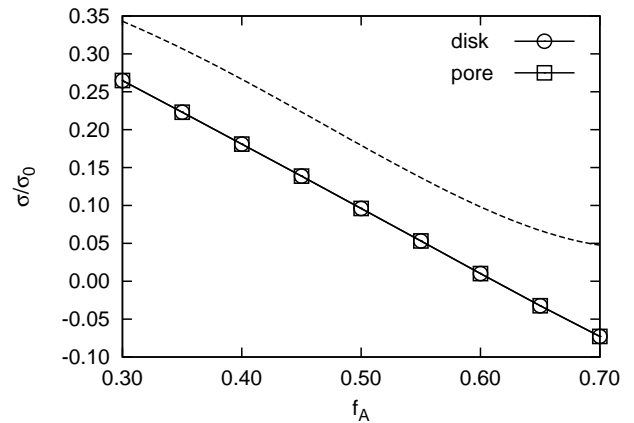}
  \caption{The dimensionless line tension $\sigma/\sigma_0$, obtained
    from the disk and pore geometries, versus hydrophilic fraction $f_A$.
    The dashed line shows the prediction from the coupled monolayer
    approximation (\ref{eq:couple_line}). }
  \label{fig:gamma}
\end{figure}

The calculated line tension as a function of the  hydrophilic volume fraction is shown in Fig.~\ref{fig:gamma}.
The line tensions obtained from the disk and pore geometries are almost identical,
reflecting the fact that when the radius $R$ is large in comparison to the bilayer thickness, there is no difference between the edge of a disk and the edge of a pore.
As seen in Fig.~\ref{fig:gamma}, the line tension decreases from positive to negative in an approximately linear way as a function of the hydrophilic volume fraction.
For most lipids, such as DOPC or DOPE, their hydrophilic fractions are less than 0.50, corresponding to a large positive line tension.
The results for the line tension imply that a closed membrane in the form of vesicles is the preferred structure in such cases.
On the other hand, the line tension of membranes formed from lipids with large hydrophilic heads can even become negative, implying that it is possible to use these big-headed lipids as edge stabilizers.
This result is also supported by simulations \cite{deJoannis2006} and experiments \cite{Karatekin2003}.
In Ref.~\cite{deJoannis2006}, it is found that the line tension of the bilayer depends linearly on the lipid tail length, which corresponds to the hydrophobic fraction $1-f_A$, and adding short-tail lipids (with large $f_A$) can be used to decrease the line tension of a membrane composed of a mixture of short- and long-tail lipids.

The line tension for the disk/pore structure can also be estimated from the monolayer properties if we perceive the edge as a folded monolayer.
In Appendix~\ref{app:line}, an analytic expression of the line tension is obtained for the two geometries,
\begin{equation}
  \label{eq:couple_line}
  \sigma = \pi \bar{\kappa}_M \frac{1 - 4\bar{c}_0 \delta }{2 \delta}.
\end{equation}
The prediction from this expression is plotted in Fig.~\ref{fig:gamma} as dashed line.
The analytic result correctly predicts that the line tension is a decreasing function of the hydrophilic volume fraction, but it overestimates the value of the
line tension.
In this calculation, we used the value $\delta$ for a planar bilayer membrane, while we can see from the profiles (insets of Fig.~\ref{fig:line}), that the radius near the edge is slightly larger than the planar bilayer.
Also, near the edge the monolayer is strongly curved and higher-order contribution to the bending energy become important, as demonstrated in the previous section. 
Furthermore, the simple analysis in Appendix~\ref{app:line} is unable to consider the contribution due to the derivative terms because of the discontinuity
at the junction of flat and curved membrane surface.
Given all these possible corrections, the qualitative agreement between SCFT results and the approximate theory (\ref{eq:couple_line}) is remarkable and indicates that the monolayer model does capture the physics of the edge elastic
properties.

\section{Summary}
\label{sec:summary}

To summarize, we have systematically investigated the elastic properties of a bilayer membrane, including the line tension of the edge, the bending modulus, the Gaussian modulus, and the fourth-order moduli, by applying the self-consistent field theory to a molecular model where the membrane is composed of amphiphilic
chains dissolved in hydrophilic solvent molecules.
The free energy of membranes subject to different geometric constraints has been
calculated.
The elastic properties were extracted from the SCFT free energy of the membrane in five geometries: the planar, the cylindrical, the spherical, the disk, and the pore geometry.
In particular, SCFT studies of a disk-like membrane and a pore in a planar membrane were used to extract the line tension of the membrane edge.
We have explored the dependence of the elastic properties on the microscopic characteristics of the amphiphiles as specified by the hydrophilic volume fraction $f_A$.
Three notable results are:
(i) As the hydrophilic volume fraction increases, the line tension decreases linearly from a positive value to zero and even to a negative value, indicating that big-headed lipids can be used as edge stabilizers.
(ii) The Gaussian modulus is a monotonically decreasing function of the hydrophilic volume fraction, whereas the bending modulus is a concave function with a maximum around $f_A=0.50$.
The ratio of the Gaussian modulus and the bending modulus changes from 1 to -2 as $f_A$ is increased from 0.30 to 0.70.
(iii) The quadratic Helfrich model is accurate at small membrane curvatures, but it becomes inaccurate when the radius of curvature is comparable to the membrane thickness.
Therefore, higher-order curvature moduli have been introduced and calculated.

The elastic properties of a bilayer membrane could also be analyzed within a two-monolayer approximation, which is capable of predicting the properties of bilayers from the monolayer properties.
Physical insights can be obtained from the relation between the elastic properties of the bilayer and the monolayer.
For example, the bending modulus of bilayers is roughly twice that of monolayers,
while the Gaussian modulus of a bilayer is a combined effect of the bending modulus, Gaussian modulus and spontaneous curvature of a monolayer.
The line tension of a disk/pore structure largely originates from the bending energy (dictated by the monolayer bending modulus) of the monolayer that forms the bilayer edge.

This work focuses on the influence of the amphiphilic architecture on the properties of single-component bilayer membranes.
It should be noticed that other factors such as the interaction strength
between different molecules can also be important.
For example, the Flory-Huggins parameter determines the scale of the curvature moduli to a large extent, because our results are scaled by the interfacial tension between the A/B homopolymers.
The present model also has a number of limitations concerning the higher-order contributions to the elastic free energy. 
Ideally, the fourth-order terms in the free energy expression (\ref{eq:Helfrich4}) also include terms related to the derivative of the curvatures. 
For the planar bilayers, cylinders and spheres, the curvatures are uniform thus the derivative terms vanish.
For the disk/pore geometries, their contribution will change the effective line tension.
Furthermore, the geometries considered in this work only allow us to obtain two linear combinations of the three fourth-order moduli. 
To extract these moduli separately, one can consider the cylinders with an ellipse cross-section or an ellipsoid in three-dimension \cite{Manyuhina2007}.

In many biological membranes, the composition of the inner and outer membrane leaflets are different. 
The asymmetric composition arises naturally in cell membranes when the lipids are generated and initially inserted into the inner leaflet, and the asymmetry often persists over long periods of time because the spontaneous flip-flop of lipids between two leaflets is extremely slow \cite{Sanyal2009}.
The current SCFT model in general deals with thermodynamical equilibrium thus it is challenging to treat dynamic processes, but one may develop clever constraint schemes to enforce the asymmetric composition.
Another possible cause of the asymmetry is that the fluids on two sides of the membrane have different chemical compositions.
This can be simulated in the present model by using homopolymers interacting differently with the copolymers.
Furthermore, multi-component bilayer membranes exhibit interesting features
\cite{Gutlederer2009, HuJinglei2011, LiJianfeng2013}, for example, lipids that are capable of stabilizing the edge tend to aggregate on the edge, resulting in a smaller line tension.
Local curvature effects may also reduce the line tension of phase separated lipid domains in multi-component membranes, or stabilize microphase separated ``raft'' structures \cite{Meinhardt2013}.
The current model can be extended to multi-component membranes by introducing two or more copolymers, even with different architectures.
The study of the elastic properties of multi-component bilayer membranes
with the SCFT will be an appealing and exciting subject of future work.

 \begin{acknowledgments}
We are grateful to Marcus M\"uller and Ashkan Dehghan for valuable discussions.
We acknowledge support from the National Basic Research Program of China (Grant No. 2011CB605700), the National Natural Science Foundation of China (Grants No. 20874019, 20990231, and 21104010), the Natural Sciences and Engineering Research Council (NSERC) of Canada through Discovery and CREATE program, and the Deutsche Forschungsgemeinschaft (DFG) through SFB 625.
This work was made possible by the facilities of the Shared Hierarchical Academic Research Computing Network (SHARCNET:www.sharcnet.ca) and Compute/Calcul Canada, and JGU Mainz (MOGON).
 \end{acknowledgments}

\appendix

\section{Two-Monolayer: Bending and Gaussian Moduli}
\label{app:moduli}

Let us assume that the bilayer is composed of two identical
monolayer leaflets, and each monolayer has the same spontaneous
curvature $\bar{c}_0$, bending modulus $\bar{\kappa}_M$ and Gaussian
modulus $\bar{\kappa}_G$. We will use symbols with top bar for the
monolayer and symbols without for the bilayer. The bilayer is
described by the area $A$ at its midplane, and the mean and Gaussian
curvature of the bilayer midplane are denoted by $M=(c_1+c_2)/2$ and
$G=c_1 c_2$, where $c_1$ and $c_2$ are principal curvatures. The
areas at the neutral surfaces of the outer and inner monolayers,
$A_{\rm out}$ and $A_{\rm in}$, are related to that of the bilayer
midplane by \cite{doCarmo, Petrov1984, Siegel2004, Marsh2006}
\begin{eqnarray}
  \label{eq:A_out}
  A_{\rm out} &=& A ( 1 + 2 M \delta + G \delta^2), \\
  \label{eq:A_in}
  A_{\rm in}  &=& A ( 1 - 2 M \delta + G \delta^2),
\end{eqnarray}
where $\delta$ is the separation between the monolayer neutral
surface to the bilayer midplane. The mean and Gaussian curvatures at
the neutral surfaces of the two monolayers are
\begin{eqnarray}
  \label{eq:M_out}
  M_{\rm out} &=& (M+G\delta) \frac{A}{A_{\rm out}}, \\
  \label{eq:G_out}
  G_{\rm out} &=& G \frac{A}{A_{\rm out}}, \\
  \label{eq:M_in}
  M_{\rm in} &=& - (M-G\delta) \frac{A}{A_{\rm in}}, \\
  \label{eq:G_in}
  G_{\rm in} &=& G \frac{A}{A_{\rm in}}.
\end{eqnarray}

The Helfrich free energy density for the monolayers are
\begin{eqnarray}
  \label{eq:f_out}
  f_{\rm out} &=& 2 \bar{\kappa}_M(M_{\rm out}-\bar{c}_0)^2 + \bar{\kappa}_G G_{\rm out}, \\
  \label{eq:f_in}
  f_{\rm in} &=& 2 \bar{\kappa}_M(M_{\rm in}-\bar{c}_0)^2 + \bar{\kappa}_G G_{\rm out}.
\end{eqnarray}
The free energy of the bilayer results from a summation of the free
energies of the outer and inner monolayer,
\begin{eqnarray}
  \label{eq:fA1}
  f A &=& f_{\rm out} A_{\rm out} + f_{\rm in} A_{\rm in} \\
  &=& \Big[ 4 \bar{\kappa}_M M^2 + 2[\bar{\kappa}_G - 4 \bar{\kappa}_M \bar{c}_0 \delta (1-\frac{1}{2} \bar{c}_0\delta) ] G \Big] A, \nonumber
\end{eqnarray}
where we have neglected the constant term and terms that are of
higher than second order in the curvatures.

The free energy of the bilayer membrane can be written in terms of
the bilayer properties
\begin{equation}
  \label{eq:fA2}
  f A = \Big[ 2 \kappa_M (M-c_0)^2 + \kappa_G G \Big] A.
\end{equation}
From Eqs.\ (\ref{eq:fA1}) and (\ref{eq:fA2}), we immediately obtain
the relations between the bilayer and monolayer properties:
\begin{eqnarray}
  c_0 &=& 0, \\
  \kappa_M &=& 2 \bar{\kappa}_M, \\
  \kappa_G &=& 2 \big[\bar{\kappa}_G - 4 \bar{\kappa}_M \bar{c}_0 \delta (1-\frac{1}{2} \bar{c}_0 \delta) \big].
\end{eqnarray}

\section{Two-Monolayer: Edge Line Tension}
\label{app:line}

\begin{figure}[htp]
  \includegraphics[width=1.0\columnwidth]{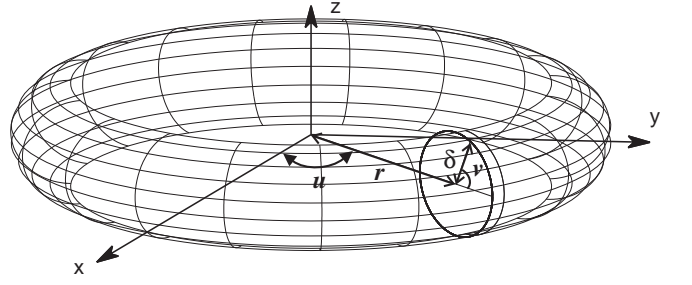}
  \caption{A sketch of the surface near the disk edge.}
  \label{fig:torus}
\end{figure}

The surface of a bilayer membrane disk near the edge can be regarded
as the outer part of a torus \cite{Safran,Wohlert2006}. The torus
surface can be parametrized using two angle $u$ and $v$ (see Fig.
\ref{fig:torus}),
\begin{equation}
  \mathbf{r}(u,v) = \begin{bmatrix}
    (r+\delta \cos v) \cos u \\
    (r+\delta \cos v) \sin u \\
    \delta \sin v
  \end{bmatrix}
\end{equation}
where $r$ is the radius of the torus center line and the thickness
of the disk is $2\delta$. For the outer side of the torus, the range
of the two angles are $u\in [0,2\pi]$ and $v\in[-\pi/2,\pi/2]$.

The two principal curvatures of this surface are
\begin{eqnarray}
  c_1 &=& \frac{1}{\delta} , \\
  c_2 &=& \frac{\cos v}{r+ \delta \cos v}.
\end{eqnarray}
The mean and Gaussian curvatures are
\begin{eqnarray}
  M &=& \frac{ 2\delta\cos v+r }{2\delta(r+\delta\cos v)} ,\\
  G &=& \frac{ \cos v }{ \delta(r+\delta\cos v) } .
\end{eqnarray}
One more quantity we need is the area element $\ud A$,
\begin{equation}
  \ud A = \sqrt{g} \ud u \ud v = \delta(r+\delta\cos v)\ud u \ud v,
\end{equation}
where $g$ is the surface metric.

The bending free energy of the disk edge is then
\begin{eqnarray}
  F^D &=& \int \big[ 2\bar{\kappa}_M (M^2 -2 M \bar{c}_0) + \bar{\kappa}_G G \big] \ud A \nonumber \\
  &=& 2\pi \int_{-\pi/2}^{\pi/2} \big[ 2\bar{\kappa}_M (M^2 - 2M\bar{c}_0) + \bar{\kappa}_G G \big] \sqrt{g} \ud v. \nonumber
\end{eqnarray}
The contribution from the Gaussian curvature can be easily evaluated
\begin{equation}
  \label{eq:F_G}
  \frac{F_G}{2\pi} = 2 \bar{\kappa}_G \int_{-\pi/2}^{\pi/2} \cos v \ud v = 2 \bar{\kappa}_G .
\end{equation}
The calculation for the mean curvature is slightly more complicated
\begin{eqnarray}
  \frac{F_M}{2\pi} &=& - 4 \bar{\kappa}_M + 8 \bar{\kappa}_M \bar{c}_0 \delta - 2 \pi \bar{\kappa}_M \bar{c}_0 r \nonumber \\
  &+& \pi \bar{\kappa}_M \frac{r}{d} \frac{1}{\sqrt{1-(\frac{d}{r})^2}} \Big( 1 - \frac{2}{\pi} \tan^{-1} \sqrt{ \frac{1-\frac{d}{r}}{1+\frac{d}{r}} } \Big). \nonumber
\end{eqnarray}
The last term in the above equation can be expanded up to order
$(\frac{d}{r})^2$ using
\begin{equation}
  \Big[ 1-(\frac{d}{r})^2 \Big]^{-\frac{1}{2}} = 1 + \frac{1}{2}(\frac{d}{r})^2 + \mathscr{O}((\frac{\delta}{R})^2)
\end{equation}
\begin{eqnarray}
  \tan^{-1} \sqrt{ \frac{ 1 - \frac{d}{r} }{ 1 + \frac{d}{r} } } &\approx& \tan^{-1} \Big( 1-\frac{d}{r} \Big) \\
  &=& \frac{\pi}{4} - \frac{d}{2r} - \frac{1}{4} (\frac{d}{r})^2 + \mathscr{O}((\frac{\delta}{R})^2), \nonumber
\end{eqnarray}
and the final result is
\begin{eqnarray}
  \frac{F_M}{2\pi} &=& \pi \bar{\kappa}_M \big[\frac{1}{2\delta} - 2\bar{c}_0 \big] r + \bar{\kappa}_M \big[ 3 - 8\bar{c}_0\delta \big] \nonumber \\
  \label{eq:F_M}
  &+& \bar{\kappa}_M \big[\frac{\pi}{4} - \frac{1}{2} \big] \frac{\delta}{r} + \mathscr{O}((\frac{\delta}{R})^2).
\end{eqnarray}

Note that the disk radius is defined as $R=r+\delta$. From Eqs.\
(\ref{eq:F_G}) and (\ref{eq:F_M}), we get the total energy of the
disk edge
\begin{eqnarray}
  \frac{F^D}{2\pi} &=& \pi \bar{\kappa}_M \big[\frac{1}{2\delta} - 2\bar{c}_0 \big] R \nonumber \\
  &+& \bar{\kappa}_M  \big[ (3 - \frac{\pi}{2}) + (2\pi - 8) \bar{c}_0\delta \big] \\
  &+& 2 \bar{\kappa}_G + \bar{\kappa}_M \big[\frac{\pi}{4} - \frac{1}{2} \big] \frac{\delta}{R}  + \mathscr{O}((\frac{\delta}{R})^2). \nonumber
\end{eqnarray}
We can identify the line tension as
\begin{equation}
  \sigma = \pi \bar{\kappa}_M \big[\frac{1}{2\delta} - 2\bar{c}_0 \big].
\end{equation}

For the pore structure, the edge surface can be regarded as the
inner part of the torus. A slight modification to the above
derivation is the integration range of parameter $v$, which changes
to $[\pi/2,3\pi/2]$. The final result is
\begin{eqnarray}
  \frac{F^P}{2\pi} &=& \pi \bar{\kappa}_M \big[\frac{1}{2\delta} - 2\bar{c}_0 \big] R \nonumber \\
  &+& \bar{\kappa}_M  \big[ (\frac{\pi}{2}-3) + (8 - 2\pi) \bar{c}_0\delta \big] \\
  &-& 2 \bar{\kappa}_G + \bar{\kappa}_M \big[\frac{\pi}{4} + \frac{1}{2} \big] \frac{\delta}{R}  + \mathscr{O}((\frac{\delta}{R})^2). \nonumber
\end{eqnarray}
The line tension for the pore is the same as that obtained for the
disk geometry.


\bibliography{line}


\end{document}